\documentclass[useAMS]{mn2e}

\newcommand{\be} {\begin{equation}}

\newcommand{\ee} {\end{equation}}

\newcommand{\XMM}{{\em XMM--Newton}}

\newcommand{\RXTE}{{\em R}XTE}
\newcommand{\CXO}{{\em Chandra}}
\newcommand{\bc}{\begin{center}}
\newcommand{\ec}{\end{center}}

\def\ltsima{$\; \buildrel < \over \sim \;$}
\def\lsim{\lower.5ex\hbox{\ltsima}}
\def\loe{\lower.5ex\hbox{\ltsima}}
\def\gtsima{$\; \buildrel > \over \sim \;$}
\def\gsim{\lower.5ex\hbox{\gtsima}}
\def\goe{\lower.5ex\hbox{\gtsima}}

\def \cm2{cm$^{-2}$\,}

\def\ergs {erg\,s$^{-1}$}
\def\ergscm2 {erg\,s$^{-1}$cm$^{-2}$}

\def\srca{LS\,I\,+61$^{\circ}$303}
\def\srcb{LS\,5039\,}
\def\srcc{HESS\,J0632+057\,}

\begin{document}

\title[Deep Chandra observations of \srca]{Deep \CXO\, observations of TeV binaries I: \srca}

\author[Rea et al.]{N. Rea$^{1}$\thanks{Ramon y Cajal Research Fellow; rea@ieec.uab.es.}, D.~F. Torres$^{1,2}$, M. van der Klis$^{3}$, P.~G. Jonker$^{4,5}$, M. M\'endez$^{6}$, \newauthor A. Sierpowska-Bartosik$^{7}$ \\
$^{1}$ Institut de Ci\'encies de l'Espai (ICE, IEEC--CSIC), Campus UAB, Fac. de Ci\'encies, Torre C5-parell, 2a planta, 08193 Barcelona, Spain \\
$^{2}$ Instituci\'o Catalana de Recerca i Estudis Avan\c{c}ats (ICREA),
Barcelona, Spain \\
$^{3}$ University of Amsterdam, Astronomical Institute ``Anton Pannekoek'', Postbus 94249, 1090 GE, Amsterdam, The Netherlands \\
$^{4}$ SRON-Netherlands Institute for Space Research, Sorbonnelaan 2, 3584 CA, Utrecht, the Netherlands \\
$^{5}$ Harvard-Smithsonian Center for Astrophysics, 60 Garden Street, Cambridge, MA 02138, USA \\
$^{6}$ Kapteyn Astronomical Institute, University of Groningen, PO Box 800, 9700 AV, Groningen, The Netherlands \\
$^{7}$ University of Lodz, Department of Astrophysics, Pormorska street 149/153, PL-90236, Lodz, Poland \\
}

\input psfig.sty

\pagerange{\pageref{firstpage}--\pageref{lastpage}} \pubyear{2009}

\maketitle

\label{firstpage}

\begin{abstract}

We report on a $95$\,ks \CXO\, observation of the TeV emitting High Mass X--ray Binary \srca, using the ACIS-S camera in Continuos Clocking mode to search for a possible X-ray pulsar in this system. The observation was performed while the compact object was passing from phase 0.94 to 0.98 in its orbit around the Be companion star (hence close to the apastron passage). We did not find any periodic or quasi-periodic signal (at this orbital phase) in a frequency range  of $0.005-175$\, Hz. We derived an average pulsed fraction 3$\sigma$ upper limit for the presence of a periodic signal of $\lsim$10\% (although this limit is strongly dependent on the frequency and the energy band), the deepest limit ever reached for this object. Furthermore, the source appears highly variable in flux and spectrum even in this very small orbital phase range, in particular we detect two flares, lasting thousands of seconds, with a very hard X-ray spectrum with respect to the average source spectral distribution. The X-ray pulsed fraction limits we derived are lower than the pulsed fraction of any isolated rotational-powered pulsar, in particular having a TeV counterpart. In this scenario most of the X-ray emission of \srca\, should necessarily come from the interwind or inner-pulsar wind zone shock rather than from the magnetosphere of the putative pulsar.  On the other hand, very low X-ray pulsed fractions are not unseen in accreting neutron star systems, although we cannot at all exclude the black hole nature of the hosted compact object, a pulsar with a beam pointing away from our line of sight or spinning faster than $\sim5.6$\,ms, nor that pulsations might have a transient appearance in only a small fraction of the orbit. Furthermore, we did not find evidence for the previously suggested extended X-ray emission.

\end{abstract}

\begin{keywords}
X-ray: binaries -- (stars:) binaries: individual: \srca
\end{keywords}

\section{Introduction}

High Mass X-ray binaries (HMXBs) are relatively young ($< 10^7$~year) 
systems composed of a
massive OB-type star and a compact object, either a neutron star or a
black hole. These systems are generally bright X-rays emitters, due to
matter from the OB star accreting onto the compact object. The
majority of the HMXBs are Be/X-ray binaries, and some tens of them 
are O main-sequence or supergiant X-ray binary systems. Accretion
takes place through either the strong stellar wind of the optical
companion or Roche-lobe overflow, where matter flows via the inner
Lagrangian point  and an accretion disk to the compact object (see Frank, King \& Raine 1992). A powerful X-ray source
(L$_{\rm X}\sim10^{35}-10^{38}$\ergs ) is produced, which ionizes
almost the whole stellar wind.  The
X-ray continuum spectra of HMXBs are often described by a power law
with photon index $\alpha \sim 1-2$ (modified at higher energies by an
exponential cutoff). A spectrum of this form can be produced by
inverse Compton scattering of soft X-rays by hot electrons in the
accretion column near the compact object, and a part of this emission
is scattered by the stellar wind of the massive companion. This
might sometimes result in a further non-thermal spectral component, but with a
different absorption column depending on the orbital phase of the
system. Furthermore, in some HMXBs a soft excess at $\sim 0.1-2$ keV
is detected, very common in systems hosting pulsars (Hickox
et~al~2004).

On top of the continuum model, spectral lines are often present in these systems, 
neutral and ionized, such as Fe, Si, Mg, Ar, N, Ca, mainly produced in the stellar wind or in the
accretion disk (if any) illuminated by the strong X-ray emission of
the compact object (see e.g. Cottam et al. 2001; van der Meer et al. 2005).


\begin{figure}
\centerline{\psfig{figure=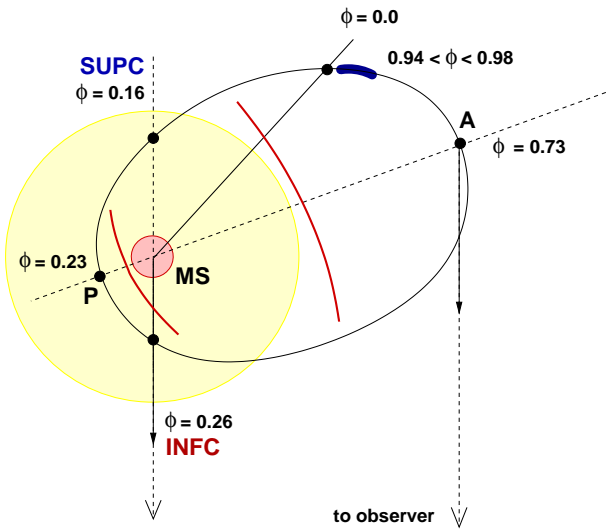,width=8cm}}
\caption{\srca's geometry considering the orbital solution of Casares et al. (2005). The phases for Inferior
conjunction (INFC), Superior conjunction (SUPC), periastron (P), and apastron (A) are marked, and the inclination is not taken into account in the plot. The orbit (black solid line), the massive star (MS; in orange) and the equatorial Be disk (in yellow) are roughly to scale,  assuming a disk radius 7 times larger than the stellar radius. The blue thick line indicates the orbital phases spanned by our \CXO\, observation (adapted from Sierpowska-Bartosik \& Torres 2010).}
\label{orbit}
\end{figure}


To date there are only four or five binary systems known to be emitting up to TeV energies, 
all of them of the HMXB class: PSR\,B1259--63
(Aharonian et al.~2005a), \srca\, (Albert et al.~2006), \srcb\,
(Aharonian et al.~2005b,~2006), Cyg\,X-1 (Albert et al.~2007) and possibly the
newly suggested candidate \srcc\, (Aharonian et al. 2007, Hinton et
al. 2009).  These sources are rather peculiar with respect to the HMXB class. 
The first identified TeV binary system was PSR\,B1259--63,
a 3.4~year period binary hosting a 48\,ms radio pulsar in an eccentric
orbit around a Be-star (Johnston et al.~1992). The TeV emission from
this object is thought to be associated with the radio pulsar wind and
its interaction with the radiation field and material around the
Be-star, and shows variable un-pulsed X-ray (Chernyakova et al. 2006, 2009) and 
radio emission (Johnston et al. 1999, 2005), in particular close to the periastron 
passage due to the interaction between the pulsar and the Be disk.  PSR\,B1259--63's X-ray emission is well described by an absorbed power-law (N$_{\rm H}=5\times10^{21}$\cm2; $\Gamma\sim$1.2-1.8), with an unabsorbed 1--10\,keV flux varying between 0.1--3$\times10^{-11}$\ergscm2 (Chernyakova et al. 2006, 2009).

\srca\, and \srcb\, are both much closer binaries, with
orbital periods of 26.5 and 3.9~days, respectively, hosting a very
massive star (B and O types) and a compact object, the nature of which 
is still unknown for both sources (see Fig.\,\ref{orbit} for a schematic view of \srca's orbit). Furthermore, \srca\, and \srcb\, share the
quality of i) being the only two known $\gamma$-ray binaries 
detected in the GeV band (Hartman et al.~1999; Abdo et al. 2009a, 2009b), ii) they
both show periodic TeV emission modulated by their orbital motion
(Aharonian et al. 2006; Albert et al. 2008, 2009), iii) variable X-ray
emission (Paredes et al.~2007; Sidoli et al.~2006; Bosch-Ramon et
al.~2005, 2007) correlated with the orbital motion, and iv) they have a variable radio counterpart (see also below). Their X-ray emission is highly variable on a ks timescale, and is for both objects characterized by an absorbed power-law spectrum (\srca: N$_{\rm H}=6\times10^{21}$\cm2 and $\Gamma\sim$1.4-1.9; \srcb:  N$_{\rm H}=7\times10^{21}$\cm2 and $\Gamma\sim$1.4-1.6), with a 1--10\,keV unabsorbed flux varying between $0.4-1.5$ and $0.5-1.2\times10^{-11}$\ergscm2 , for \srca\, and \srcb, respectively (Sidoli et al. 2006; Esposito et al. 2008; Kishishita et al. 2009).

As the nature of the compact object is unknown in these two
systems, it is not clear if their emission (especially in the TeV band) 
is due to a relativistic outflow from a rotational powered pulsar (as the case of
PSR\,B1259--63) or accretion onto a black hole or a neutron star
which might e.g. accelerate particles through a relativistic
jet. Recently, a Soft Gamma Repeater (SGR)-like burst detected from the direction
of \srca\, raised the possibility of this system hosting a magnetar,
although this is so far a very controversial hypothesis (see de
Pasquale et al.~2008; Barthelmy et al~2008; Dubus \& Giebels~2008; Rea
\& Torres~2008, Ray et al.~2008; Munoz-Arjonilla et al.~2008). In
the accretion scenario context, the recent evidence for TeV emission from the black hole
binary Cyg\,X-1, is very exciting (Albert et al.~2007).  However, the
very high energy phenomenology of Cyg\,X-1 appears different from that
of these other sources: it has been in fact detected in TeV just once,
and during a flaring state. The three other sources, instead, present
persistent TeV emission modulated by the orbital motion, and which
dominates their radiative outputs. For the new candidate TeV binary
\srcc, there is still only very little known (e.g. no orbital period
detected yet) to make any solid comparison (Aharonian et al. 2007,
Hinton et al. 2009; Maier et al. 2009).


\begin{figure*}
\centerline{\psfig{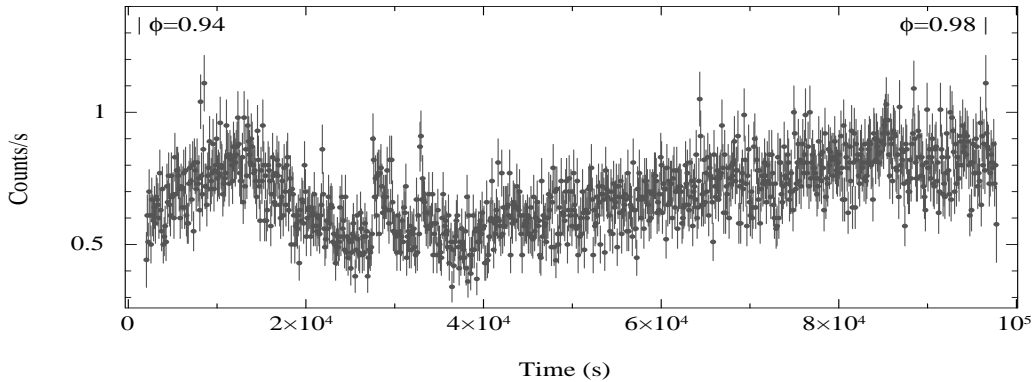}}
\caption{\CXO\, background-subtracted lightcurve of \srca\, binned at 100\,s, and in the 0.3--10\,keV energy range. The orbital phase $\phi$ is also reported. }
\label{lcurve}
\end{figure*}


For \srca , extended radio emitting structures at angular extensions
of 0.01-0.05\arcsec \, have been reported (Massi et
al.~2001,~2004). This discovery has earlier supported its microquasar
interpretation, as a claim for a jet-like feature. However, a
subsequent large VLBI campaign (Dhawan et al.~2006), discovered rapid
changes during the orbit, in the orientation of what seems to be a
cometary tail consistent with being the result of a pulsar wind more
than a jet--like structure. In fact, if due to a jet, the changing
morphology of the radio emission along the orbit would require a
highly unstable jet, and the shape of this structure is not expected
to be reproduced orbit after orbit as current results indicate (Albert
et al.~2008). Similarly, the discovery of a jet-like radio structure
in \srcb, initially prompted a microquasar interpretation (Paredes et al. 2000).
However, Dubus (2006) showed that this extended radio emission
could also be interpreted as the result of a pulsar wind, with no need
to invoke a radio jet. Furthermore, recent VLBA observations showed
that a microquasar scenario cannot easily explain the observed
morphology changes of this radio extended feature (Rib\'o et
al.~2008). 

At this point, one of the long-standing open questions of high energy
astrophysics is: are these two peculiar binary systems accelerating charged
particles until TeV energies, e.g. through the powerful jets of an accreting neutron star or black hole, as it might be the case of the TeV flares from Cyg X-1? Or is the
TeV emission of these binaries instead related to a young and energetic rotational powered-pulsar as the case of PSR\,B1259--63? To try to address this question we performed deep \CXO\, observations of \srca\, (this paper) and \srcb (Rea et al. 2010, in preparation) in search for pulsed X-ray emission with very low pulsed fraction (e.g. because contaminated by the wind-shock emission), which due to the higher background of previous X-ray observations might have been missed. In \S\ref{data} we report on the \srca\, observation and analysis, and we present our results in \S\ref{results}, followed by a discussion (\S\ref{discussion}).


\begin{figure}
\centerline{\psfig{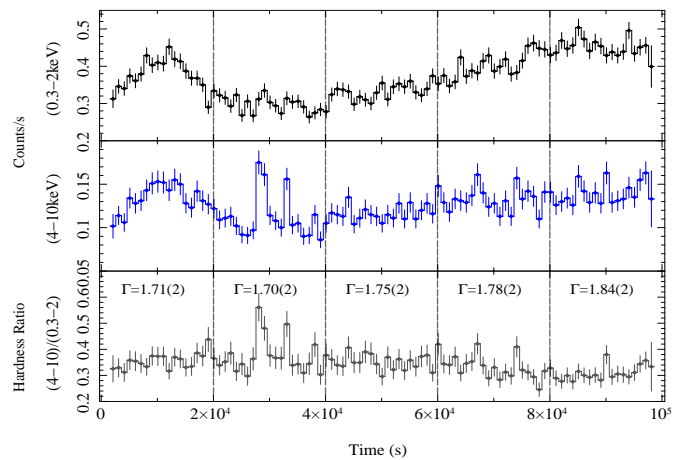}}
\caption{\srca\, lightcurve in the 0.3--2\,keV and 4--10\,keV energy bands (top and middle rows, respectively), and their ratio (bottom row), all reported with a 1\,ks bin time. Vertical lines separate the 5 intervals used in the analysis (of which we report the photon index ($\Gamma$) values). }
\label{hardness}
\end{figure}


\section{Observation and Data analysis}
\label{data}
The Advanced CCD Imaging Spectrometer (ACIS) camera on board of \CXO\,
observed \srca\, on 2008 November 14th (start time 11:09:58 UT; Obs-ID 10052) for an
exposure time of 95.668\,ks in Continuos Clocking (CC) mode
(FAINT). The CC observing mode provides a time resolution of 2.85 ms
and imaging along a single direction. The source was positioned in the
back-illuminated ACIS-S3 CCD at the nominal target position. Standard
processing of the data was performed by the {\em Chandra  X-ray Center}  to
level 1 and level 2 (processing software DS ver. 7.6.11.9). The data
were reprocessed using the CIAO software (ver. 4.1) and the \CXO\,
calibration files (CALDB ver. 4.2.0). \\
\indent 
Since in the CC mode the events are tagged with the times of the frame
store, we corrected the times for the variable delay due to the
spacecraft dithering and telescope flexure, starting from level 1 data
and assuming that all photons were originally detected at the target
position. Furthermore, data were filtered to exclude hot pixels, bad columns, and possible afterglow
events (residual charge from the interaction of a cosmic ray in the
CCD). Photon arrival times are in TDB and were referred to the
barycenter of the Solar System using the JPL-DE405 ephemeris. \\
\indent
In order to carry out a timing analysis, we extracted the events in
the 0.3--10\,keV energy range from a region of 5$\times$5 pixels around the
source position (RA 02:40:31.670, Dec $+$61:13:45.11) to reduce the background contamination, while the source 
spectrum was extracted from a rectangular region of 5$\times$25 pixels around the source position, and
the background was taken independently from a source-free region in
the same chip. We extracted the response matrix files (RMFs) and
ancillary response files (ARFs), first creating a weighted image,
re-binning by a factor of 8, then using it to build the RMF file
using the {\tt mkacisrmf} tool, with an energy grid ranging from 0.3
to 10 keV in 5 eV increments. Using this RMF and the aspect histogram
created with the aspect solution for this observation ({\tt asphist}),
we generated the appropriate ARF file for the source position. The source ACIS-S count rate in the 0.3--10\,keV energy band was  $0.648\pm0.003$\,counts\,s$^{-1}$ (the lightcurve is shown in Fig.\,\ref{lcurve} and Fig.\,\ref{hardness}).


\begin{figure*}
\hbox{
\vbox{
\psfig{figure=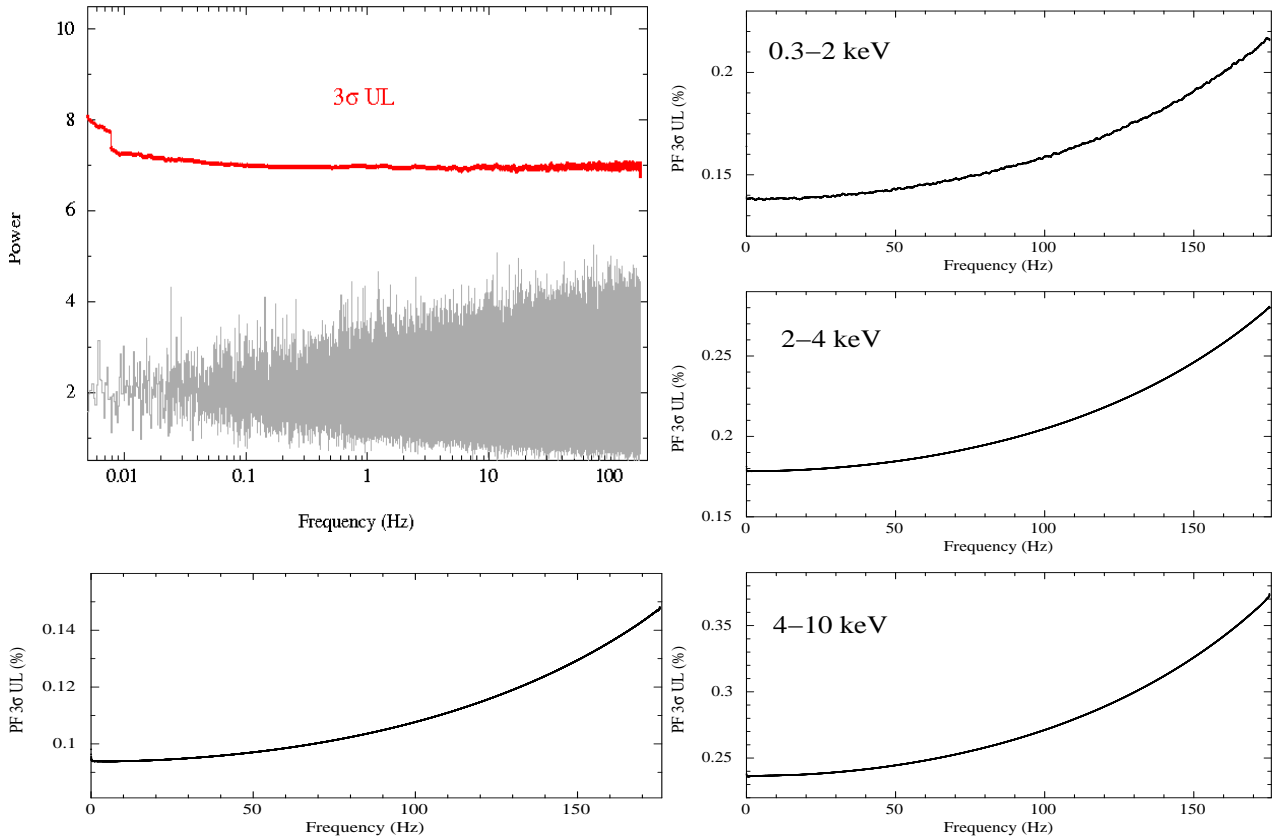,width=8.5cm,height=7cm,angle=180}
\vspace{0.5cm}
\psfig{figure=pf_000285_last.ps,width=8.7cm,height=3.5cm,angle=270}}
\vbox{
\psfig{figure=pf_000285_03_2keV_last.ps,width=8cm,height=3.5cm,angle=270}
\vspace{0.2cm}
\psfig{figure=pf_000285_2_4keV_last.ps,width=8cm,height=3.5cm,angle=270}
\vspace{0.2cm}
\psfig{figure=pf_000285_4_10keV_last.ps,width=8cm,height=3.5cm,angle=270}}}
\caption{{\em Left panels:} Top row: power spectra of \srca\, in the 0.3--10\,keV energy band binned at 0.00285\,s with superimposed the 3$\sigma$ upper limits on the detection of a periodic signal on top of the Poisson noise (we used here logarithmic scales to better show the low frequency range). Bottom row: pulsed fraction 3$\sigma$  upper limits on the semi-amplitude of a sinusoidal signal. {\em Right panels:}  pulsed fraction 3$\sigma$ upper limits versus frequency in three different energy ranges (from top to bottom: 0.3--2\,keV, 2--4\,keV, 4--10\,keV), calculated on the power spectrum binned at 0.00285\,s (left panel) in the 0.005--175\, Hz frequency range (see \S\ref{pulsation} for further details).}
\label{dps}
\end{figure*}


\section{Results}
\label{results}

Using the orbital ephemeris of Casares et al. (2005), namely an
orbital period of 26.4960$\pm$0.0028\,days and the zero-phase assumed
at T$_0$=2,443,366.775\,JD (defined as the radio phase; Gregory
2002), we calculated that the \CXO\, observation was performed while
the neutron star was passing from phase 0.94 (at the start of the
observation) to phase 0.98 (when it ends). Even if considering other orbital solutions (Grundstrom et al. 2006; Aragona et al. 2009), this corresponds for all assumed solutions  to a part of the orbit soon after the apastron passage (see Fig.\,\ref{orbit}).

\subsection{Timing analysis: pulsation search and flares}
\label{pulsation}

We searched for periodic and quasi periodic signals with unprecedented sensitivity, thanks to the extremely low ACIS-S background (see for previous limits: Sidoli et al. 2006; Harrison et al. 2002\footnote{Note that the \RXTE\, pulsed fraction limits reported by Harrison et al. 2002 ($\sim6$\%) are calculated from the total count rate without correcting for the cosmic and instrumental background, which if corrected increases substantially the upper limit on the detectable pulsed fraction to $\sim32$\%.}; Torres, Sierpowska-Bartosik \& Rea 2010). Given the length of our observation (95\,ks) and the timing resolution of the CC-mode (0.00285\,s), we could in principle search for periodic signals in the 10$^{-5}$ -- 175\,Hz frequency range (namely between the Nyquist frequency  and half of the frequency resolution of our \CXO\, data; van der Klis 1989), however, our data were sensitive only in the 0.005 -- 175\,Hz frequency range, which we will then use all over this study. We studied the source power spectra performing Fast Fourier Transforms (FFTs; see Fig.\,\ref{dps}) using the {\tt Xronos} analysis software, and we did not find any periodic or quasi-periodic signal neither considering the total exposure time nor in each of the five time intervals shown in Fig.\,\ref{hardness}.  Furthermore, we performed the same search dividing the data in three energy bands (0.3--2, 2--4, and 4--10\,keV), again not finding any significant signal. 
\indent
In particular, for computing reasons, we performed an average over 9 FFTs with a bin time of 0.00285\,s (Fig.\,\ref{dps} top-left panel), resulting in 4,194,304 bins ($\sim12$\,ks) for each of the 9 averaged power spectra (the averaged power spectrum had a $\chi_{\nu}^{2}$ distribution with 18 d.o.f.). For the power spectra produced for the three different energy bands, given the lower number of counts, we could use a single FFT (their power spectrum had then a $\chi_{\nu}^{2}$ distribution with 2 d.o.f.) . We took into account the number of bins searched, and the different d.o.f. of the noise power distribution (Vaughan et al. 1994; Israel \& Stella 1996) in calculating the 3$\sigma$ detection upper limits  reported in Fig.\,\ref{dps} . At very low frequencies, most probably due to the overall variability and flares present in the X-ray lightcurve (see Fig.\,\ref{lcurve}), there is evidence for the onset of a  red-noise component. We accounted for the red-noise component in the calculation of the detection significance by using a smoothing window technique (see Israel \& Stella 1996 for further details).  \\
\indent
We computed the 3$\sigma$ upper limits on the sinusoid semi-amplitude pulsed fraction ($PF$), computed according to Vaughan et al. (1994) and Israel \& Stella (1996), which ranges in the 0.3--10\,keV energy band between $PF<$7--15\% (see Fig.\,\ref{dps}  bottom-left panel).  In  Fig.\,\ref{dps} (right column) we calculate the same pulsed fraction limits as a function of the energy band. It is straightforward that the lower number of counts present in the energy-selected lightcurves causes the energy-dependent $PF$ limits to be slightly larger than those derived using the whole dataset.   Note that given the long orbital period of this binary system ($\sim26.5$\,days; Casares et al. 2005), we could reliably ignore the effect of such orbit during our pulsar searches, since our exposure time is much less than 10\% of orbital period (Vaughan et al. 1994). \\
\indent
We looked for the presence of SGR--like bursts by binning the counts in intervals of 0.1\,s and searching for excesses above a count threshold corresponding to a chance occurrence of 0.1\% (taking into account the total number of bins), but we did not find any significant short burst. On the other hand, on top of a slow variability of a factor of $\sim 2$, we significantly detect two flares lasting $\sim$3 and 1.5\,ks (not unseen in this source: Sidoli et al. 2006; Paredes et al. 2007; Esposito et al. 2007; Smith et al. 2009) during the minimum of the lightcurve,  with a rise-time of $\sim100$\,s, and peaking after about 28 and 33\,ks from the observation start (see Fig.\,\ref{lcurve} and \ref{hardness}).

\subsection{Spatial analysis: search for extended emission}
\label{spatial}

We studied the radial profile of \srca, using the CC mode one-dimensional imaging capability, to search for any evidence for extended X-ray emission (see also Paredes et al. 2007). We first generated an image of the one-dimensional strip in the 0.3--10\,keV band, and then subtracted the background count rate to remove instrumental and cosmic X-ray background. We then produced a one-dimensional surface brightness distribution using the same method as for a two-dimensional radial profile, namely extracting photons from 125 annuli centered on the source position and of 2 pixel wide ($\sim$0.99\arcsec). We then simulated a {\em Chart/MARX} Point Spread Function (PSF) and extracted photons from the same annular regions. We did not find evidence for a significant extended emission or structures within 30\arcsec\, around our target (see Fig.\,\ref{profile}). As a further  check, given the possible uncertainties of the one-dimensional CC-mode PSF, we extracted with the same method a similar profile from a source observed in the same mode, and known to be point-like, namely the AMCVn RX\,J0806+1527 (Obs-ID 4523; Israel et al. 2003). This source has a spectrum similar to our source, this is why we chose it as a good candidate for the PSF comparison. As shown in the inset of Fig.\,\ref{profile}, we do not find any excess in the ratio between the surface brightness of \srca\, and RX\,J0806+1527, but only a constant flux difference of $\sim31$, which corresponds to the different brightness of these sources.

\subsection{Spectral analysis: continuum modeling and search for spectral lines}
\label{spectral}

The total spectrum, and the spectra relative to the 5 intervals shown in Fig.\,\ref{hardness}, were binned such to have at least 60 counts per bin and not to over-sample the instrument spectral resolution by more than a factor of 3. We fit the total spectrum with an absorbed power-law (typical for this source; see e.g. Bosch-Ramon et al. 2007) using the XSPEC software, which gives a reduced $\chi_{\nu}^2=1.33$ for 173 degree of freedom (d.o.f.). The $\chi_{\nu}^2$ was not completely satisfactorily, we then studied the spectrum diving the total observation in 5 time intervals (see Fig.\,\ref{hardness} and \ref{spectra}) to investigate whether this was due to a significant spectral variability within the observation. This was indeed the case. We fit together the spectra of the 5 time-intervals with an absorbed power-law function finding a $\chi_{\nu}^2=1.01$ for 598 d.o.f, with the absorption value tied to be the same for all 5 spectra and the slope and normalization of the power-laws free to vary. We find an absorption value of N$_{\rm H}=(6.10\pm0.08)\times10^{21}$\cm2, and a power-law index which varied from 1.70 to 1.83 depending on the time-interval (see Fig.\, \ref{hardness}; all errors in the spectral values are reported at 1$\sigma$ confidence level). The source 0.5--10\,keV absorbed flux varied from the first to the fifth time interval (from left to right in Fig.\,\ref{hardness}): $8.5\pm0.2$, $6.6\pm0.2$, $7.2\pm0.1$, $8.5\pm0.2$ and $9.0\pm0.1 \times10^{-12}$\ergscm2 (corresponding to unabsorbed fluxes of $1.2\pm0.1$, $0.9\pm0.1$, $1.0\pm0.1$, $1.2\pm0.1$ and $1.3\pm0.1 \times10^{-11}$\ergscm2 ). \\  We also tried a fit using a {\tt pegpwr} model instead of the canonical {\tt powerlaw} model. The difference between the two is  that in the former the normalization is calculated on an energy range given by hand (we assumed 0.3-10\,keV), while in the latter case it is calculated at 1\,keV by default, and this can sometime cause a spurious $\Gamma$-normalization dependence. Using a {\tt pegpwr}  we found  $\Gamma$ and flux values consistent within 1$\sigma$ with what previously reported for a simple power-law fit. \\
\indent We also tried to leave the absorption value free to vary among the 5 spectra. We found an absorption value varying between 5.8(2)-6.2(2)$\times10^{21}$\cm2, with the same photon index or flux variation as the previous modeling. We then conclude that we do not detect any significant absorption variability within the sensitivity of our data.
\indent We do not find any evidence for absorption or emission features in the source averaged and time-resolved spectra. We derived upper limits as a function of the line energy and width by adding Gaussian lines to the continuum model both in absorption and in emission. For the total spectrum, the 3$\sigma$ upper limit on the equivalent width of the presence of lines range between $50-550$\,eV depending on the energy (between 0.3--10\,keV) and assumed line width ($\sigma_{\rm E} = 0.05-0.5$\,keV). As a simpler visual plot, we report in Fig.\,\ref{spectra} (right panel) the 1$\sigma$ upper limit on a detectable flux variation with respect to our continuum model as a function of the energy band, which corresponds to the detectability of spectral lines as broad as the energy-dependent instrument resolution on top of the continuum flux.

\section{Discussion}
\label{discussion}

Thanks to this new \CXO\, observation (see \S\ref{data}) we inferred the deepest upper limits to date on the presence of X-ray pulsations from the TeV Be binary \srca\, ($PF \lsim10$\%; see \S\ref{pulsation}, and Fig.\,\ref{dps}), while the compact object was close to the apastron of its orbit around the Be companion (see Fig.\,\ref{orbit}). Furthermore, we observed a strong flux variability during this observation (eventhough it was spanning a relatively small orbital phase in the 0.94-0.98 range), in particular \srca\, emitted two ks-timescale flares during our observation, with a rather hard spectrum with respect to the source  persistent emission (see Fig.\,\ref{hardness} and \S\ref{spectral}).  The accurate \CXO\, spatial resolution allowed us to study the X-ray spatial profile (although with the one-dimensional CC-mode capability; see \S\ref{spatial}), which we found consistent with the instrumental PSF.

\subsection{X-ray pulsed fractions of rotational-powered and accreting pulsars, comparison with our current limits}
\label{pulsars}

The detection of pulsations is the only unambiguous tracer for a secure determination of the pulsar nature of the compact objects hosted in \srcb\, and \srca . Deep searches for pulsations have been performed in the radio band at several frequencies, with the hope of detecting a fast spinning radio pulsar as in the case of the other TeV binary PSR\,B1259--63 (Johnston et al. 1999, 2005). However, no radio pulsations have been detected so far from any of these two sources. This is not surprising; in fact the strong massive companion winds might have prevented radio pulsations to be detected because of the strong free-free absorption all over the orbits and/or the large and highly variable Dispersion Measure (DM) induced by the wind. In particular, in that respect it is crucial to note that at periastron PSR\,B1259--63 does not show radio pulsations, and its periastron (given the large orbit, 3.4 year period) has about the same dimension of the major axis of the orbit of \srca\,  (26 days period) and it is way larger than LS\,5039's obit (4 days period), making very plausible the difficulty in detecting any radio pulsed emission from these closer binaries.

On the other hand, searches for pulsations in the X-ray band have many more chances of success than in the radio band. In fact: the X-ray pulsar beam is usually larger than the radio one, and the strong companion wind does not influence much the X--ray pulsed emission if present.

The deep $PF$ upper limits we derived here (see \S\ref{results}, and Fig.\,\ref{dps}) can be used to compare the X-ray emission of this TeV binary with that of isolated rotational-powered pulsars.  X--ray pulsations from rotational powered pulsars have been detected in tens of objects, of which only fourteen are to date also TeV emitters (see e.g. Mattana et al. 2009), due to the presence of an energetic Pulsar Wind Nebula (PWN). All these TeV emitting pulsars known to date have high rotational power  ($3\times10^{35}<\dot{E}<5\times10^{38}$\,erg~s$^{-1}$), fast pulsations ($ 33 < P <  326$\,ms), young characteristic ages ($0.7 < \tau_{\rm c} < 90$~kyr), rather high magnetic fields ($10^{12} < B < 5\times10^{13}$\,Gauss), and X-ray pulsed fractions $PF\gsim18$\%.


\begin{figure}
\centerline{\psfig{figure=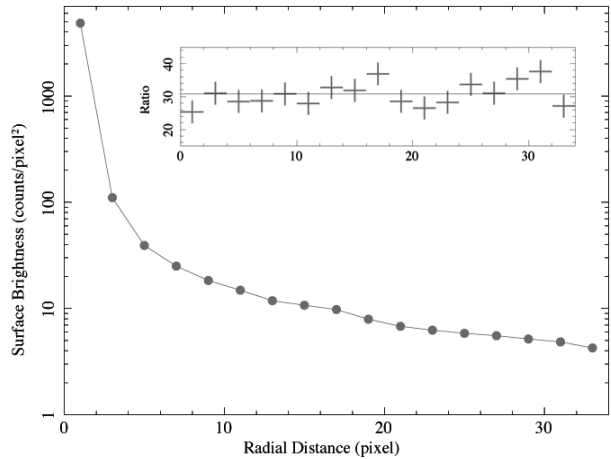,width=8cm,height=6cm,angle=270}}
\caption{Surface brightness of \srca. 1 pixel corresponds to 0.495\arcsec . The inset represents the ratio between \srca\, and RX\,J0806+1527's surface brightnesses.}
\label{profile}
\end{figure}


The X-ray upper limits we derived here, namely $PF\sim$10\% (see Fig.\,\ref{dps}) for spin periods as fast as $\sim5.6$\,ms (much faster than the fastest rotational powered pulsar known to be a TeV emitter), clearly indicates that if the compact object hosted in \srca\, is indeed a rotational-powered pulsar, its X-ray emission cannot be dominated by synchrotron emission from the pulsar magnetosphere alone, which might at most account for $\lsim$10\% of the emission, while the shock between the pulsar and the massive companion star winds is responsible for the remaining X-ray emission of the system. 

In fact in this pulsar scenario for \srca, radio, X-ray and gamma-ray emission are (at least in part) thought to come from the region where the pulsar wind interacts with the wind of the Be companion star and/or (particularly, the gamma-rays) from within the pulsar wind zone. This mostly depends on where the high-energy electrons generating the synchrotron and inverse Compton emission originate, the opacities therein, and the geometry of the orbit (Sierpowska-Bartosik \& Torres 2008) . 

In this respect it should be mentioned that the TeV binary system hosting PSR\,B1259--63, does not show X-ray pulsations (Chernyakova et al. 2006) during its periastron passage, with a limiting pulsed fraction (derived as described in \S\ref{pulsation}) of $\sim 15$\% (more details on that will be presented in the second paper of this series)\footnote{Note that Chernyakova et al. (2006) reports a limiting pulsed fraction of 2\%, however these authors did not report the exact definition they used for pulsed fraction, the energy range, and background subtraction method they used. Possible differences in these assumptions might be the cause of such discrepancy.}. Also for this source, the X-ray emission might in fact be dominated by the inter-wind and/or pulsar-wind zones rather than being due to emission from the pulsar magnetosphere.


\begin{figure*}
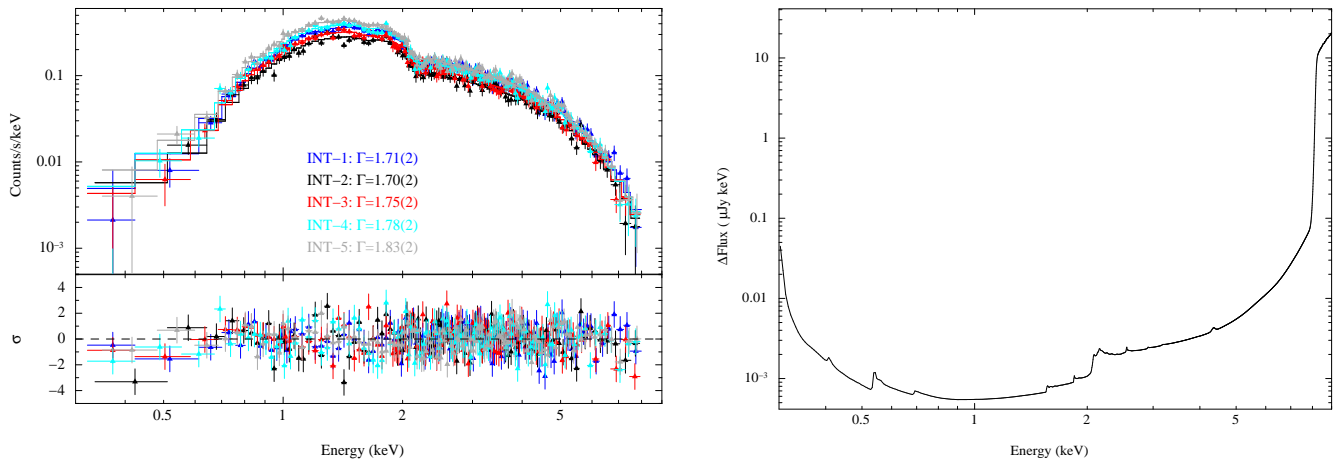

\hbox{
\psfig{figure=spectral_evolution_2.ps,width=8.7cm,height=6cm,angle=270}
\hspace{0.5cm}
\psfig{figure=insensitivity_plot.ps,width=8.2cm,height=6cm,angle=270}}
\caption{{\em Left panel}:  spectra of the 5 time intervals fitted with an absorbed power-law function (see Fig.\,\ref{hardness} and \S\ref{spectral} for details). {\em Right panel}: 1$\sigma$ sensitivity for the presence of spectral lines (defined as the 1$\sigma$ limit for the detection of flux variability as a function of energy).}
\label{spectra}
\end{figure*}


However, caution should be used here: in this line of reasoning we are neglecting the possibility of  a pulsar with a beam pointing away from our line of sight or spinning faster than $\sim5.6$\,ms, and also that strong pulsations might have a transient appearance in only a small fraction of the orbital motion, and in such small fraction being instead dominated by the pulsar magnetosphere.  On the other hand, if we assume that a putative transient pulsed emission is diluted in part of the orbit by Compton scattering, from the rise times of the flares $t \sim R_{\rm C} \ \tau /c \sim 100$\,s (see Fig.\,\ref{lcurve}; $c$ is the velocity of light), and assuming an optical depth $\tau  >>1$, we can derive that the Compton scattering sphere radius should be $R_{\rm C}<<3\times10^{12}$\,cm, which is about half of semi-major axis of \srca's orbit. In this respect, our observations were performed in a part of the orbit (soon after the apastron passage; see Fig.\,\ref{orbit}), where the detection of transient pulsations should have been more favorable,  and the Compton scattering highly reduced with respect to other parts of the orbit closer to the massive companion.

Considering instead the accreting neutron star scenario (in the black hole case we obviously do not expect pulsations at all hence we will not discuss this possibility which is of course open), low pulsed fractions are not unexpected. In particular, in many cases the pulsed X-ray emission coming from the accretion column might represent only a small fraction of the total X-ray emission which can also have a dominant  un-pulsed component coming from the accretion disk and/or the neutron star surface.

\subsection{Flares, and flux/spectral variability on ks timescales}
\label{flares}

A large variety of flares on several timescales (from seconds to days) have been observed to date from both neutron star and black hole binary systems. While in the accretion scenario  flares are easily explainable as changes in the accretion rate from the companion star, or accretion disk, towards the compact object, this kind of X-ray variability has never been observed in isolated rotation powered-pulsars\footnote{Note that magnetars do show X-ray flaring activity on several timescales, but in fact they are not rotational-powered while their emission (flaring and persistent) is due to the presence and instability of their ultra-strong magnetic fields (see Mereghetti 2008 for a recent review).}.

In the rotational-powered pulsar scenario, the ks-variability we observe here indicates that the interaction of the putative pulsar and the stellar wind change at  distance scales which are smaller than the size of the binary orbit; put otherwise, that the  mechanism by which the variability is generated has a shorter-than-orbital timescale. To interpret the short timescale variability (of the order of ks, and in the form of hard flares) seen in our \CXO\,  data (also seen earlier in \XMM\, observations by Sidoli et al. (2006) and Chernyakova et al. (2006)), it seems reasonable to invoke clumpiness of the stellar wind (see ideas by Neronov \& Chernyakova 2007, and Zdziarski, Neronov \& Chernyakova 2008).

If one assumes that the power output in the pulsar wind is constant in ks timescales (which may not be necessarily true, due to inhomogeneities in the pulsar wind zone, and if so, ks variability could be related to it), the distance scale related to the fast variability has to be associated with the pulsar wind interactions with clumps of the Be stellar wind. In the latter case, the size, mass, and number density of the clumps imprint their signal on the X-ray emission.  The distance scale relevant for this fast variability can be estimated as:
 
$$R_{\rm w}\sim v\Delta t \sim 10^{11}\left ( \frac{v}{10^7\mbox{ cm/s}}
\right) \left(\frac{\Delta t}{10\mbox{ ks}}
\right) \mbox{ cm} $$

\noindent
where $v$ is typical velocity scale (e.g. the orbital velocity of the 
pulsar, or the speed of the stellar wind).

The stellar wind of Be stars, like the one present in \srca, is typically
assumed to have two components, one related with a polar contribution, and the other,
with an equatorial disk (e.g. Waters et al. 1988). 
The polar wind is radiatively driven (e.g. Castor \& Lamers 1979). The equatorial wind is instead assumed to have an homogeneous velocity law  
$
V_{\rm eq}(r) = V_o^{eq} \left( {r}/{R_{\rm s}}\right)^{m}
\label{eq:vd}
$
where $m \sim 1.25$, $R_{\rm s}$ is the stellar radius, and $V_0^{eq} = 5\,\rm km\, s^{-1}$.  These authors further assume  that the terminal velocity of the equatorial wind is $V_{\infty}^{eq} \sim \rm few\, 100\, km\, s^{-1}$, so that it is about a factor of at least 10 times smaller than the terminal velocity of the polar component, $V_{\infty}^{polar}$. Clumpiness in this wind can make the region of collision between the pulsar and stellar outflows to loose coherence, leading to the disappearance of a regular bow-shaped surface. 
If so happens,  the escape timescale of the
pulsar wind mixed with the stellar wind can be written as

$$t_{\rm esc}\sim D/v_{\rm wind}\sim
100    \left(   \frac{10^{12}\mbox{ cm}}{D}\right)  \left(\frac{10^7\mbox{
cm/s}}{v_{\rm wind}}\right)   \mbox{ ks} 
$$

\noindent
(slowed down from the case in which  
electrons move along the shock with the plasma velocity $c/3$, the  drift velocity in the direction parallel to the shock surface), and where $v_{\rm wind}$ can take typical $V_{\infty}^{eq}$ or $V_{\infty}^{polar}$ values.

Even with this timescale, clumps can cool electrons via inverse Compton. This statement can be entertained after comparing the cooling timescale of electrons of about 10 MeV, which would be responsible via inverse Compton emission of X-ray photons between 1 and 10\,keV.
Assuming that electrons diffuse in disordered
magnetic field in the clump, one can estimate the Bohm-diffusion-regime escape time as:

$$
t_{\rm diff}\simeq 3\times 10^6\left(\frac{R_{\rm w}}{10^{11}\mbox{ cm}}\right)^2
\left( \frac{B}{1\mbox{ G}}\right)
\left(\frac{10\mbox{ MeV}}{E}\right) 
\mbox{ s}
$$

\noindent
thus, clumps with such magnetic fields can retain electrons over significant timescales. 
The looses by synchrotron and inverse Compton are

$$
t_{\rm synch}  \simeq  3\times 10^5\left( \frac{1\mbox{ G}}{B}\right)^2
\left(\frac{10\mbox{ MeV}}{E}\right) \mbox{ s}, \nonumber $$

$$ t_{\rm IC} \simeq 10^4 \left( \frac{L_*}{10^{38}\mbox{ erg/s}}\right)
\left(\frac{D}{10^{12}\mbox{ cm}}\right)^2
\left(\frac{10\mbox{ MeV}}{E}\right) \mbox{ s}
$$

\noindent
with $L_*$ being the luminosity of the star. For nominal values of the parameters, then, an inverse Compton cooling is possible.

Using the expression for the inverse Compton and synchrotron
timescales one finds that if the magnetic field does not significantly
rise from its pulsar wind nebula value of $\sim $ 1 G towards the
center of the system, inverse Compton losses dominate at the inner
region. However, synchrotron emission quickly dominate the farther out
of the system electrons are, and can contribute to the overall X-ray
emission detected, given the nebula an onion-like structure where
losses dominance change. Depending on the location of the
inhomogeneity and value of magnetic field, relative contributions of
inverse Compton and synchrotron X-rays can vary.

Finally, if clumps can be described by variations in density, and if the typical density of a clump in the wind is $n_{\rm clump} \sim 10^{10}$ cm$^{-3}$ and its size is $R_{\rm clump} \sim 10^{11}$ cm, then $\delta N_{\rm H} \sim n_{\rm clump} \times R_{\rm clump} \sim 10^{21}$\cm2 . We did not find significant  $N_{\rm H}$ variability within our data, however the strong dependence between $N_{\rm H}$, $\Gamma$ and the power-law normalization, intrinsinc to the power-law modeling, might have hidden this variability in an increase of the hardness of the emission, as seen in our Fig.\,\ref{hardness} and \ref{spectra}  (see also Smith et al. 2009).

\subsection{Limits on the presence of diffuse emission}
\label{diffuse}

 Studying the one-dimentional \CXO\, PSF, we did not find evidence for X-ray diffuse emission (see Fig.\,\ref{profile} and \S\ref{spatial}). For a quantitative comparison of our limits with the hint for diffuse emission observed  by Paredes et al. (2007; see their \S2.3), we followed their analysis extracting the counts in an annular region of 5\arcsec-12.5\arcsec radii. Assuming an absorbed power law  spectrum with N$_{\rm H}=6.1\times10^{21}$\cm2 and $\Gamma=2$, we inferred a 3$\sigma$ limiting unabsorbed flux for the presence of X-ray  diffuse emission of $\sim1\times10^{-14}$\ergscm2  in the 0.3--10\,keV energy range\footnote{Note that to infer the diffuse flux, Paredes et al. (2007) assumed the spectrum of their \CXO\, observation, $\Gamma=1.25$, which was highly affected by pile-up. We then prefer to assume $\Gamma=2$, the typical value for a PWNe.}. Using the same assumptions we made, the flux of the diffuse emission proposed by Paredes et al. (2007) would be 
 $\sim2\times10^{-14}$\ergscm2 , well above our detection limit. 
  
We caveat  that if a putative extended emission (Paredes et al. 2007) shows  a long term variability, our non detection can be due to a different observing epoch with respect to the original \CXO\, observation which hinted at the presence of diffuse emission.

\section{Summary}

With the \CXO\, observation we report here, we inferred the deepest upper limits to date on the presence of X-ray pulsations from the TeV Be binary \srca\, ($PF \lsim10$\%; with a strong energy and frequency dependence, see Fig.\,\ref{dps}), while the compact object was close to the apastron of its orbit around the Be companion (see Fig.\,\ref{orbit}). These $PF$ limits are deeper than the pulsed fraction measured from any isolated rotational-powered pulsar, in particular from those emitting in the TeV range, while they are well in line with what is observed from accreting neutron star binaries (although we do not see clear signs of accretion, as e.g. spectral lines).  Hence, if hosting a rotational-powered pulsar, \srca's X-ray emission cannot be driven by the same processes as for a similar isolated case (with the caveats discussed in \S\ref{pulsars} though), while indeed a major contribution to its X-ray emission should come from the shock between the pulsar and Be star winds. This is probably also the case for PSR\,B1259--63. The ks-timescale flares we saw in this \CXO\, observation of \srca\, show a harder spectrum with respect to the source emission, and can be explained in an accretion scenario by variability in the accretion rate, while in the rotational-powered pulsar one, as the interaction between the pulsar wind and clumps in the Be wind.

\vspace{0.5cm}

This research has made use of data from the Chandra X-ray Observatory and software provided by the Chandra X-ray Center. NR is supported by a Ramon y Cajal Research Fellowship to CSIC, and thanks G.L. Israel for useful discussion, advices concerning  the timing analysis, and for allowing the use of his {\tt DPS} software. We thank the anonymous referee for his/her very useful suggestions, and V. Bosch-Ramon and M. Rib\'o for comments. This work has been supported by grants AYA2009-07391 and SGR2009-811.

\label{lastpage}

\end{document}